\documentclass[prb,showpacs,twocolumn,floatfix,superscriptaddress]{revtex4}
\usepackage{epsfig}

\begin{document}

\title{Conductance oscillations of a spin-orbit stripe with polarized contacts}
\author{M. M. Gelabert}
\affiliation{Departament de F\'{\i}sica, Universitat de les Illes Balears,
E-07122  Palma de Mallorca, Spain}
\author{Ll.\ Serra}
\affiliation{Departament de F\'{\i}sica, Universitat de les Illes Balears,
E-07122  Palma de Mallorca, Spain}
\affiliation{ Institut de F\'{\i}sica Interdisciplinar i de Sistemes Complexos
IFISC (CSIC-UIB), E-07122 Palma de Mallorca, Spain.}

\date{March 30, 2010}

\begin{abstract}
We investigate the linear conductance of a stripe of spin-orbit interaction
in a 2D electron gas; that is, a 
2D region of length $\ell$ along the 
transport direction and infinite in the transverse one in which a spin-orbit 
interaction of Rashba type is present.
Polarization in the contacts 
is described by means of Zeeman fields.
Our model predicts two types of conductance oscillations:
Ramsauer oscillations in the minority spin transmission, when both 
spins can propagate, and Fano oscillations when only one
spin propagates. 
The latter are due to the spin-orbit coupling with quasibound 
states of the non propagating spin.
In the case of polarized contacts in antiparallel configuration Fano-like oscillations
of the conductance 
are still made possible by the spin orbit coupling, even though no spin component 
is bound by the contacts.
To describe these behaviors we propose a 
simplified model based on an ansatz wave function.
In general, we find that the contribution for vanishing transverse momentum dominates 
and defines the conductance oscillations.
Regarding the oscillations with Rashba coupling intensity, our model confirms the 
spin transistor behavior, but only for high degrees of polarization. 
Including a position dependent effective mass yields additional 
oscillations due to the mass jumps at the interfaces. 
\end{abstract}
\pacs{71.70.Ej, 72.25.Dc, 73.63.Nm}
\maketitle

\section{Introduction}

Spintronics attempts to manipulate  the electron spin, alone or in combination 
with the electron charge,
to tune the current in a device or as the bit of information.\cite{fab07}
This novel technology may lead to faster responses and lower power consumptions
as compared to the more conventional electronics. 
A promising approach to spintronics is the control of the spin-orbit interaction,
particularly, of the Rashba interaction.\cite{Rashba60}
This is a spin-orbit coupling due to the lack of inversion symmetry in semiconductor 
heterostructures, such as those based on InAs or GaAs semiconductors, for which 
tunability using gate contacts has been demonstrated.\cite{nit97,eng97}
Exploiting this tunability, 
Datta and Das proposed the spin field-effect transistor,\cite{Dat90}
a ballistic channel between two ferromagnetic leads 
where current can be manipulated by changing the Rashba strength via an external gate.
Despite its simplicity, the limitations of the physical system\cite{schmidt} 
and of the ballistic transport\cite{schliemann}
proved to be long-time obstacles to realizing this device.
A recent experiment, however,  has overcome these problems demonstrating the 
feasibility of the Datta-Das transistor.\cite{Koo09}

The Rashba Hamiltonian is composed of two spin-dependent terms;
one conserves the longitudinal momentum while the other
couples longitudinal and transverse momenta. They
are usually called precession and mixing
terms, respectively.
The mixing term is not present in ideal 1D Datta-Das transistors.
More realistic quasi-1D systems are usually considered including 
a confinement in the transverse direction to transport. The quantum
wire with homogeneous spin orbit interaction has been discussed, for instance, in 
Refs.\ \onlinecite{gov04,ser05,ber07}.
In a similar quantum wire configuration but with the Rashba coupling restricted to a 
finite region of the wire, 
recent works showed the importance of
the mixing term in the modulation of
the conductance.\cite{she04,zhan05,san06,lop07,san08} 
These modulations are examples of Fano resonances
due to the coupling with quasi bound states. 
Since this behavior is caused by the spin-orbit coupling alone  
it has been called the Fano-Rashba effect.\cite{san06}

The lateral dimension of the transport channel in the experiments by 
Koo {\em et al.}\cite{Koo09} was several microns, which indicates a
high degree of 2D character, thus deviating from the 1D or quasi-1D regimes.
The quasi-1D multichannel case with polarized contacts was considered in 
Refs.\ \onlinecite{cah03,jeon06,gel10}
while the 2D system, without confinement in the transverse direction, has been also 
addressed in Refs.\ \onlinecite{mat02,kho04,pal04,baby,zai,entin}. 
In most cases sharp transitions between the contacts and the channel are assumed
and matching of the wave functions at the interfaces is the required 
condition. As an alternative, our approach assumes smooth transitions and describes 
transmission and reflection between contacts and channel from 
the numerical wave function, solution of the complete Schr\"odinger equation.
Our purpose is to provide additional insight on the origin and characteristics 
of different types of conductance oscillations.

In this work we extend the analysis of Ref.\ \onlinecite{gel10} to the case of vanishing 
transverse confinement.
We thus focus our interest in a two dimensional electron gas (2DEG) with a stripe 
of spin-orbit interaction. Polarization in the contacts is modeled
by means of effective Zeeman fields, treating the cases of parallel and antiparallel
polarizations in arbitrary directions. We assume these Zeeman fields in the contacts 
are the stray fields of nearby ferromagnets deposited on top of the 2DEG.\cite{wro01}
The direction of the Zeeman field is modeling the orientation 
of the ferromagnet for each contact. In particular,
we are interested in parallel and antiparallel orientation of the ferromagnets
in the plane of the nanostructure.
The linear conductance of our model system displays two kinds of oscillations: Ramsauer 
oscillations when the two spins are propagating
and Fano oscillations when only one spin can propagate while the other one is evanescent.
Looking at the evolution of the linear conductance with the 
Rashba coupling intensity, 
our model confirms the Datta-Das oscillation of the conductance, but only for high 
degrees of polarization in the contacts.

This paper is organized as follows.
In Sec.\ II we describe the physical system and the model for the current and 
linear conductance.
Section III discusses the dependence of the conductance on energy while
the Datta-Das transistor configuration is studied in Sec.\ IV.
In Sec.\ V we study how the results are affected by a position dependent effective mass.
Finally, Sec.\ VI presents our conclusions.

\section{Physical system and model}

\begin{figure}[t]
\centerline{
\epsfig{file=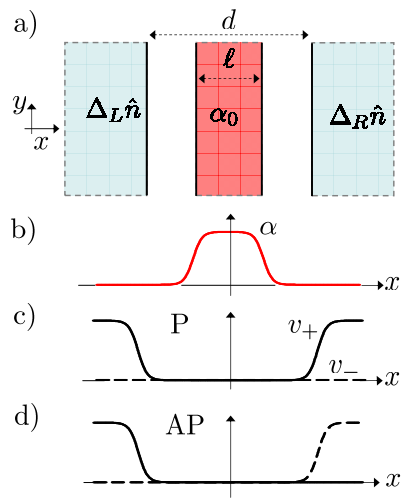,angle=0,width=0.30\textwidth,clip}}
\caption{(Color online)
Sketch of the physical system (a) and of the spatial variation
of Rashba intensity $\alpha(x)$ (b) and of the spin-dependent 
potentials $v_\pm(x)$ (c,d). See Sec.\ II.}
\label{fig1}
\end{figure}

We consider a semiconductor 2DEG in the $xy$-plane with a 
region of Rashba spin-orbit interaction shaped like 
an infinite stripe of width $\ell$ oriented along $y$.  
Figure \ref{fig1}a shows a sketch of the physical system.
Transport is along $x$ and the asymptotic leads (contacts) are assumed to 
be spin polarized along a given direction $\hat{n}$.
The system Hamiltonian reads
\begin{eqnarray}
\mathcal{H}&=&-\frac{\hbar^2}{2m_0}\left(\frac{d^2}{dx^2}
+\frac{d^2}{dy^2}\right)\nonumber\\
&+&\Delta(x)\,\hat{n}\cdot\vec{\sigma}+|\Delta(x)|+\mathcal{H}_R,
\label{eqH}
\end{eqnarray}
where $\mathcal{H}_R$ is the Rashba Hamiltonian,
\begin{equation}
\label{eqHR}
\mathcal{H}_R=\frac{1}{\hbar}\left( \alpha(x)p_y\sigma_x
-\frac{1}{2}\{ \alpha(x),p_x\}\sigma_y \right)\; .
\end{equation}
Polarized leads in the direction of $\hat{n}$ are described by
means of a Zeeman field $\Delta(x)$ that couples to the spin vector
$\vec{\sigma}$. A positive scalar potential $|\Delta(x)|$ is also 
introduced in order to align the majority spin potentials
in the contacts with the potential bottom of the central region. 
This eliminates the effect of a
potential mismatch for this spin component and, in practice,
it would correspond to use a potential gating of the central region.
In Eq.\ (\ref{eqH}) the functions determining the Hamiltonian
are $\Delta(x)$ and $\alpha(x)$. These quantities take a constant 
value in the three parts of our system: left contact (L), central 
region and right contact (R), and they vary smoothly, described by a Fermi-type
function, at the interfaces. See Appendix \ref{appA} for the precise definitions.

We denote by $m_0$ the conduction-band effective mass of the 
semiconductor and by $\alpha_0$ the Rashba intensity
of the central region. The Zeeman field in contact
$c$, where $c=L,R$, is denoted by $\Delta_c$, respectively. 
The case of parallel polarized
contacts (P) corresponds to $\Delta_L=\Delta_R\equiv\Delta_0$, while the
case of antiparallel polarizations (AP) corresponds to 
$\Delta_L=-\Delta_R\equiv\Delta_0$, where $\Delta_0$ is half
of the absolute Zeeman splitting. For simplicity, $\Delta_0$ is assumed
equal in both contacts.
We use the notation
$\hat{n}$P and $\hat{n}$AP to indicate parallel and antiparallel 
configurations along a certain direction $\hat{n}$.
Figure \ref{fig1}b shows the variation 
of the Rashba intensity $\alpha(x)$. It also shows the potentials $v_s$,
for $s=\pm$ spins, defined as
\begin{equation}
v_s(x)=s\Delta(x)+|\Delta(x)|\; .
\end{equation}
Notice that in the P configuration the $s=-$ spin sees no potential at all
while $s=+$ is confined by a potential well of width $d$. On the contrary, in 
the AP configuration both spins feel a potential step, but in 
opposite contacts. As we will discuss below, these differences in potential 
landscape for $+$ and $-$ spins greatly influence the transport properties
of the stripe with polarized contacts.

Typical values of the spin-orbit intensity for InAs based semiconductors
can be tuned around $\alpha_0\approx 10\,{\rm meV}\,{\rm nm}$, with about one
order of magnitude range. Assuming a system length of $\ell\approx 1\, \mu{\rm m}$
and a Zeeman splitting of $\Delta_0\approx 0.3\; {\rm meV}$ this implies that,
in adimensional units, one has $\alpha_0\approx 0.3 \sqrt{\hbar^2 \Delta_0/m_0}$
and $\ell\approx 10\, \sqrt{\hbar^2/m_0\Delta_0}$.
Having in mind these orders of magnitude we shall explore 
the variation 
with energy, for energies around the Zeeman gap, and for Rashba coupling 
intensities around 0.5 in adimensional units.

Asymptotically, in the 2D contacts, the Hamiltonian eigenfunctions
factorize as a plane 
wave times a spinor in the direction of $\hat{n}$,
\begin{equation}
\label{eq4}
\Phi_{cs}(\vec{r},\eta;\vec{\kappa}_{cs})=
\exp{(i\vec{\kappa}_{cs}\cdot\vec{r}\,)}\, \chi_s(\eta)\; ,
\end{equation}
where $c=L,R$ and $s=\pm$ are labelling contact and spin, respectively.
The wavenumber $\vec{\kappa}_{cs}\equiv(k_{cs},q_{cs})$ is composed of
the longitudinal ($k_{cs}$) and transverse ($q_{cs}$) components. Anticipating 
a result emphasized below, we note that the transverse momentum
is a good quantum number of the system Hamiltonian, Eq.\ (\ref{eqH}). 
Therefore,
$q_{cs}$ must be a characteristic of the wave function not only in
an asymptotic region $c$ and for a given spin $s$, but throughout 
the system, i.e., $q_{cs}\equiv q$.
At a given Fermi energy $E$ we then have 
\begin{equation}
\kappa_{cs}^2=
k_{cs}^2+q^2
=
\frac{2m_0}{\hbar^2}(E-s \Delta_c-|\Delta_c|).
\label{kappaF}
\end{equation}

The physically acceptable wave functions
fulfill Schr\"odinger's equation
\begin{equation}
\label{eqSch}
(\mathcal{H}-E)\Psi=0.
\end{equation}
A most general wave function can be taken as a sum on spins and over all 
transverse momenta (an integral in $q$)
\begin{equation}
\Psi(\vec{r},\eta)=
\sum_{s=\pm}\int{dq\, \psi_{qs}(x)\, e^{iqy}\,\chi_{s}(\eta)}\; ,
\end{equation}
where the unknown functions $\psi_{qs}(x)$ can be interpreted as 
the wave amplitudes in each channel given by $(qs)$.
Projecting Eq.\ (\ref{eqSch}) we obtain the channel amplitude equations
\begin{eqnarray}
&&
\left( 
-\frac{\hbar^2}{2m_0}\frac{d^2}{dx^2}+
\frac{\hbar^2q^2}{2m_0}
+v_s(x)-E 
\right)
\psi_{qs}(x)
\nonumber\\
&+&\sum_{s'=\pm} 
\left\{ \left( 
\alpha(x)q\langle s|\sigma_x|s'\rangle
+\frac{i}{2}\alpha'(x)\langle s|\sigma_y|s' \rangle 
\right)
\psi_{qs'}(x)\right. \nonumber\\
&&\hspace*{1cm}+ 
\left.
i\alpha(x)\langle s|\sigma_y|s'\rangle\frac{d}{dx}\psi_{qs'}(x)
\right\}=0.
\label{ccmeq}
\end{eqnarray}
Notice that the channel equations for different $q$'s are uncoupled due to the 
translational invariance of the system in the transverse direction. 
At a given $q$,
however, the two spin components do couple with each other due to 
the Rashba spin-orbit interaction. This coupling is described in Eq.\ (\ref{ccmeq})
by the matrix elements 
$\langle s|\sigma_{x,y}|s'\rangle$ 
which can not be diagonalized simultaneously. For any orientation of the 
spin quantization axis $\hat{n}$, therefore, there is a Rashba-induced
interference of the two spin projections. In the contacts the 
Rashba coupling vanishes and the wave function recovers the 
good spin eigenstates $\Phi_{cs}$ given in Eq.\ (\ref{eq4}).

Integration of Eq.\ (\ref{ccmeq}) determines the transmission $T_{s's}$,
which represents an electron entering the system from the left contact
with spin $s$ and going to the right lead with spin $s'$.
It also gives $T_{s's}'$, i.e., from the right contact with spin $s$ to the 
left one with spin $s'$. In terms of these transmissions 
the total current $I_x$, per unit of length in the transverse
direction $L_y$, can be obtained by adding up the contributions 
of all electrons in each contact,\cite{Ferry}
\begin{eqnarray}
\frac{I_x}{L_y} &=&
\frac{e}{(2\pi)^2}
\int_{k>0}{
d^2\!\kappa\, \frac{\hbar k}{m_0}
\sum_{ss'}{f_{Ls}^{(+)}(\vec{\kappa})\, T_{s's}(\vec{\kappa}) }
}
\nonumber\\
&+& 
\frac{e}{(2\pi)^2}
\int_{k<0}{
d^2\!\kappa\, \frac{\hbar k}{m_0}
\sum_{ss'}{ f_{Rs}^{(-)}(\vec{\kappa})\, T_{s's}'(\vec{\kappa}) }
}
\; .
\label{eqcu}
\end{eqnarray}

In Eq.\ (\ref{eqcu}), $f_{cs}^{(\pm)}(\vec\kappa)$ represents the distribution 
function of electrons
in contact $c$ with spin $s$, with the upper index indicating
right ($+$) or left ($-$) direction of motion of the corresponding electron. 
The distribution 
of electrons in each contact is given by a Fermi function, characterized by a 
given chemical potential $\mu_c$. 
In the linear response regime the bias $\delta V = \mu_R-\mu_L$ is 
very small and it is enough to retain the linear conductance
$I_x=G\delta{V}$. From Eq.\ (\ref{eqcu}) we find the 
conductance per unit of transverse length
\begin{eqnarray}
\frac{G}{L_y} &=& \frac{G_0}{4\pi}\sum_{ss'}
\left\{
\kappa_{Ls}\, 
\int_{-\pi\!/2}^{\pi\!/2}{\!\!d\theta\, |\cos\theta|\, T_{s's}(\kappa_{Ls},\theta)}
\right. 
\nonumber\\
&+&
\left.
\kappa_{Rs}
\int_{\pi\!/2}^{3\pi\!/2}{
\!\!d\theta\, |\cos\theta|\, 
T_{s's}'(\kappa_{Rs},\theta)}
\right\}\;,
\label{eqG}
\end{eqnarray}
where $G_0=e^2/h$ is the conductance quantum and $\kappa_{cs}$ is
the Fermi wavevector in contact $c$, given by Eq.\ (\ref{kappaF}) 
when $E$ is the corresponding Fermi energy.

\section{results}

\subsection{Numerical energy dependence}

\begin{figure}[t]
\centerline{
\epsfig{file=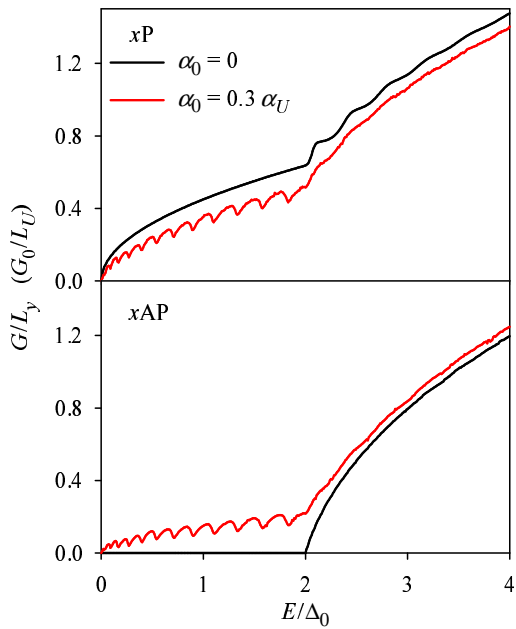,angle=0,width=0.38\textwidth,clip}
}
\caption{(Color online)
Conductance as a function of energy for polarized contacts
along $x$ in P (upper) and AP (lower) configurations. Gray (red in color) and black
curves are the results with and without Rashba stripe, respectively.
The Zeeman field parameter
$\Delta_0$ is taken as energy unit and, accordingly, $L_U=\sqrt{\hbar^2/m_0\Delta_0}$ as length unit.
The unit of Rashba coupling intensity is given by
$\alpha_U=\sqrt{\hbar^2\Delta_0/m_0}$.
Other Hamiltonian parameters: $d=20\,L_U$, $\ell=8\,L_U$, $\sigma=0.1\,L_U$.
}
\label{fig2}
\end{figure}

\begin{figure}[t]
\centerline{
\epsfig{file=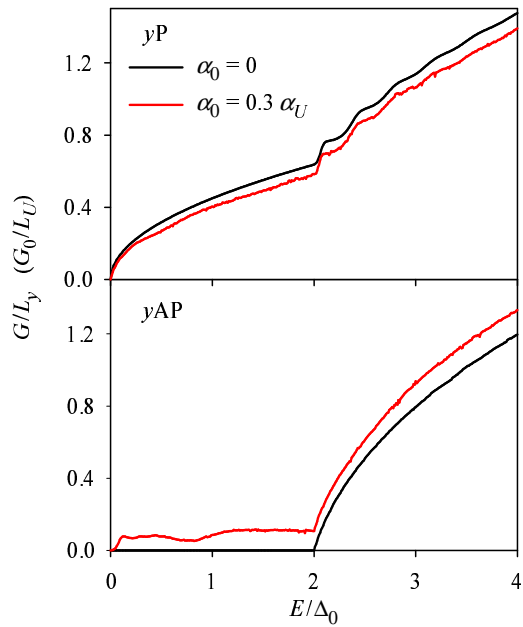,angle=0,width=0.38\textwidth,clip}
}
\caption{(Color online)
Same as Fig.\ \ref{fig2} for polarization of the contacts 
along $y$.
}
\label{fig2b}
\end{figure}

Figures \ref{fig2} and \ref{fig2b} show typical results obtained numerically 
from Eq.\ (\ref{eqG}) for $x$- and $y$-polarized contacts, respectively.
The case of polarization along $z$ is similar to the $x$ one.
For each direction ($x$ and $y$) we also compare the situation of
parallel and antiparallel Zeeman fields in the contacts.

For a vanishing Rashba field the transmissions can be obtained 
analytically. If, in addition to $\alpha_0=0$, the Zeeman fields also 
vanish ($\Delta_0=0$), the transmissions trivially become one and the exact 
conductance is then 
\begin{equation}
G=G_0\frac{2}{\pi}\,\sqrt{\frac{2m_0 E}{\hbar^2}}\; .
\label{eqsqe}
\end{equation}
When $\Delta_0\ne 0$ we have to distinguish P and AP configurations. 
The P case is characterized by a perfect transmission of the $-$ spin, while 
the $+$ spin feels the $v_+$ potential of Fig.\ \ref{fig1}. Therefore, 
its transmission switches on only when $E>2\Delta_0$. When this occurs,
the underlying potential well makes the transmission of the $+$ spin 
oscillate with energy, even with vanishing spin-orbit. 
Following Ref.\ \onlinecite{cah03},
we call
these variations Ramsauer oscillations, in analogy 
with the Ramsauer effect in electron scattering.\cite{Schiff}
The importance of these oscillations was pointed out
in Ref.\ \onlinecite{mat02}.
Notice also that with vanishing Rashba field the results for $x$ and $y$ 
orientations of the Zeeman fields are identical.

The energy $2\Delta_0$ signals the transition threshold from 
only one propagating spin when $E<2\Delta_0$, to both spins when $E>2\Delta_0$.
At $\alpha_0=0$, the P conductance below threshold is given by a 
pure square root behavior, as in Eq.\ (\ref{eqsqe}), while above threshold
it shows Ramsauer oscillations of the minority spin transmission. The 
AP transmission is exactly zero below threshold ($\alpha_0=0$) and above
it begins to increase smoothly, as expected for a spin valve. 
Note that the Ramsauer effect is not active 
in the AP configuration since the underlying potential is a step, instead of
a well.

Turning now to the spin orbit effects, the most conspicuous one is that for
$E<2\Delta_0$, when the 
contacts are fully polarized, the Rashba field
induces the appearance of oscillations
in the $x$P and $x$AP configurations (Fig.\ \ref{fig2}).
As we will discuss in detail in the next subsection, these 
oscillations are due to resonant Fano interferences 
between the propagating spin and the quasibound
states of the opposite evanescent spin. They are qualitatively similar
to the Fano-Rashba interferences discussed in 
Ref.\ \onlinecite{san06}  for quantum wires. 
Here, however,  
the quasibound states are caused by the polarized contacts and not by
the Rashba field itself.  In the $y$ orientation (Fig.\ \ref{fig2b}), the Fano 
oscillations below threshold are absent, and only
some small variation from the vanishing spin-orbit case can be seen.
In general, as shown in Figs.\ \ref{fig2} and \ref{fig2b}, for the $P$ configuration the 
results with Rashba coupling (gray-red curve) are slightly below the results
without spin-orbit (black curve); while For the AP configuration the situation
is reversed.

\begin{figure}[t]
\centerline{
\epsfig{file=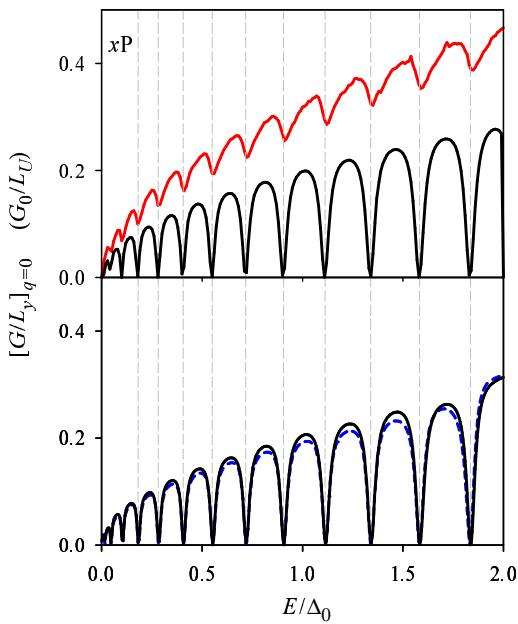,angle=0,width=0.38\textwidth,clip}
}
\caption{(Color online)
Contribution with $q=0$ to the linear conductance 
$G/L_y$. The vertical dashed lines signal the bound-state 
energies of the $v_+$ potential. 
Except for $\alpha_0=0.1\,\alpha_U$, we have used the same parameters and units of Fig.\ \ref{fig2}.
{\bf Upper panel}: $q=0$ conductance from Eq.\ (\ref{eqG}). For comparison, 
the gray (red) curve shows the full $G/L_y$ of Eq.\ (\ref{eqG}).
{\bf Lower panel}: $q=0$ conductance from the ansatz model. Solid and dashed 
lines are from Eqs.\ (\ref{fmod}) and (\ref{smod}), respectively.
}
\label{fig4}
\end{figure}

\subsection{The ansatz model}

Let us focus our attention on the oscillations that appear in the 
fully-polarized-current case, when $E<2\Delta_0$, both in parallel and
antiparallel configurations.
The upper panel of Fig.\ \ref{fig4} shows the transmission for vanishing
transverse momentum in $x$P configuration, in comparison with the total
transmission obtained by integrating over angle $\theta$ in 
Eq.\ (\ref{eqG}). We clearly see that the oscillations in $G$ 
are due to the deep minima in the transmission for $q=0$.
Besides, the position of these minima coincide with the 
energies of the bound states in the $v_+$ potential (dashed lines). 
At finite $q$'s, not shown in the figure, the transmission minima are shifted or
they can even disappear.
Physically,
we indeed expect the $q=0$ contribution to dominate the conductance since in
this case all the available energy is used in the longitudinal wavenumber.
The present transmission minima are examples of Fano resonances due to the 
interference with quasibound states. To better understand this behavior
this subsection presents a simplified model involving 
the quasibound states in an explicit way.

Assuming $q=0$ and $x$P configuration
Eq.\ (\ref{ccmeq}) transforms to
\begin{eqnarray}
\left(
-\frac{\hbar^2}{2m_0}\frac{d^2}{dx^2}-E
\right)\,
\psi_-
&=&
V_m \psi_+\, ,\label{eqpro}\\
\left(
-\frac{\hbar^2}{2m_0}\frac{d^2}{dx^2}+v_+-E
\right)\,
\psi_+&=& 
- V_m \psi_-\; ,\label{eqev}
\end{eqnarray}
where we have defined the gradient-dependent mixing potential
$V_m\equiv\alpha'(x)/2+\alpha(x)d/dx$. Equations
(\ref{eqpro}) and  (\ref{eqev}) constitute a two-channel
model, where $\psi_-$ is propagating while $\psi_+$ is evanescent, with
a localized mixing described by $V_m$. 
Similar models were obtained for impurities in quantum wires, where
semianalytical solutions were worked out using Green functions\cite{GL}
or the ansatz by Nockel and Stone.\cite{NS}  

Following Ref.\ \onlinecite{NS} let us make the following ansatz for the 
evanescent channel amplitude
\begin{equation}
\label{ans}
\psi_+(x)=\sum_{n}{A_n\,\phi_n(x)}\; ,
\end{equation}
where the $A_n$'s are constants and the
$\phi_n$'s are the bound state wave functions obtained by neglecting 
interchannel  mixing in Eq.\ (\ref{eqev})
\begin{equation}
\left(
-\frac{\hbar^2}{2m_0}\frac{d^2}{dx^2}+v_+-\varepsilon_n
\right)\,
\phi_n = 0\; .
\end{equation}
Notice that if $V_m$ is also neglected in Eq.\ (\ref{eqpro}) the propagating channel
corresponds to a free particle in 1D. Then, in terms of the free-particle Green function,  
we may write the general solution of Eq.\ (\ref{eqpro})
\begin{equation}
\label{gsol}
\psi_-=e^{ikx} + \frac{m_0}{i\hbar^2 k}\int_{-\infty}^\infty{dx'  e^{ik|x-x'|}\,[V_m\psi_+]_{x'} }\; ,
\end{equation}
where $k=\sqrt{2m_0E}/\hbar$ and $[V_m\psi_+]_{x'}$ denotes the action of the 
gradient-dependent potential on $\psi_+$ at point $x'$. 
Using now the ansatz (\ref{ans}) in Eq.\ (\ref{gsol}),
substituting in Eq.\ (\ref{eqev}) and projecting
on the set of bound states $\{\phi_n,n=1,\dots,N_b\}$ we obtain a matrix equation
for the $A_n$'s
\begin{equation}
\label{meq}
\sum_{n_2=1}^{N_b}{
\left[
(\varepsilon_{n_1}-E)
\delta_{n_1n_2} 
-
{\cal M}_{n_1n_2}
\right]
\, A_{n_2}
}
=
{\cal B}_{n_1}\; ,
\end{equation}
where 
\begin{eqnarray}
{\cal M}_{n_1n_2} &=& \frac{m_0}{i\hbar^2 k}\times \nonumber\\
&&
\!\!\!\!
\!\!\!\!
\int{
dx_1dx_2
[V_m\phi_{n_1}]_{x_1}
[V_m\phi_{n_2}]_{x_2}
e^{ik|x_1-x_2|}
}\; ,\nonumber\\
\end{eqnarray}
and
\begin{eqnarray}
{\cal B}_{n_1} &=& 
\int{dx
\,[V_m\phi_{n_1}]_x
\,e^{ikx}
}\; .
\end{eqnarray}

Taking the limit $x\to\infty$ in Eq.\ (\ref{gsol}) we find the 
amplitude of the transmitted wave 
\begin{equation}
\label{fmod}
t = 1 + \frac{m_0}{i\hbar^2 k}\sum_{n}{A_n {\cal B}_n^*}
\; ,
\end{equation}
and the corresponding transmission $T=|t|^2$. 
The solid line in Fig.\ \ref{fig4} lower panel displays numerical results
obtained by solving the matrix equation (\ref{meq}), showing clear transmission 
minima when the energy is close to a bound state. A more
explicit role of the bound states can be seen neglecting nondiagonal
terms of the matrix ${\cal M}$. In this case, the transmission amplitude
reads
\begin{equation}
\label{smod}
t=1+\frac{m_0}{i\hbar^2k}\sum_{n}{
\frac{|{\cal B}_n|^2}{\varepsilon_n-E-{\cal M}_{nn}}
}\; .
\end{equation}
When $E\approx\varepsilon_n$ the denominator in the right hand side 
of Eq.\ (\ref{smod}) reaches 
a minimum, thus yielding the mechanism by which the bound states
produce deep minima in transmission. Notice also that ${\cal M}_{nn}$ plays the
role of a complex ``self energy'' that slightly distorts the position 
of the minima. 
Nevertheless,
displacements of the dips from the bound state energies
are hardly seen for weak Rashba couplings
since ${\cal M}_{nn}\approx \alpha_0^2$.
It can also be shown that the relation 
${\rm Im}\left({\cal M}_{nn}\right)=-m_0|{\cal B}_n|^2/\hbar^2k$ is fulfilled and that
this implies an exactly vanishing conductance at the dip energies. 

In the AP configuration no potential well explicitly appears in the Hamiltonian,
as shown is the lower panel of Fig.\ \ref{fig1}. Nevertheless, the results 
of Fig.\ \ref{fig2} prove that the $x$AP configuration 
also shows clear oscillations, with conductance dips in similar 
positions to the $x$P-polarized case. We can explain this quasibound states
as a result of the combination of two effects: a) the reflection on the 
potential steps in $v_+$ and $v_-$; and b) the Rashba induced spin flip. Indeed, 
adequately combined, the reflection and the spin flip may lead to a trapped state
of the electron. Mathematically, we could describe this mechanism by transforming 
the Hamiltonian with a local spin rotation ${\cal D}(x)=e^{-i\sigma_z\phi(x)}$,
where $\phi(x)$ evolves from zero in the left contact to $\pi$ in the right one.
In the transformed problem one component is then effectively bound by the
two original potential steps. The transformation is rather cumbersome, however, 
due to the noncommutation of the kinetic term with
$\phi(x)$, in addition to the also noncommuting Pauli matrices.

In the $y$ orientation of the contacts there is no coupling between $+$ and $-$ 
spins for $q=0$, as immediately noticed from
the term $\alpha(x)q\langle s|\sigma_x|s'\rangle$ of Eq.\ (\ref{ccmeq}).
This explains why there are no clear Fano oscillations for energies below
threshold in Fig.\ \ref{fig2b}. The minor features, can be attributed to 
the Rashba-induced coupling for finite $q$'s.

\begin{figure}[t]
\centerline{
\epsfig{file=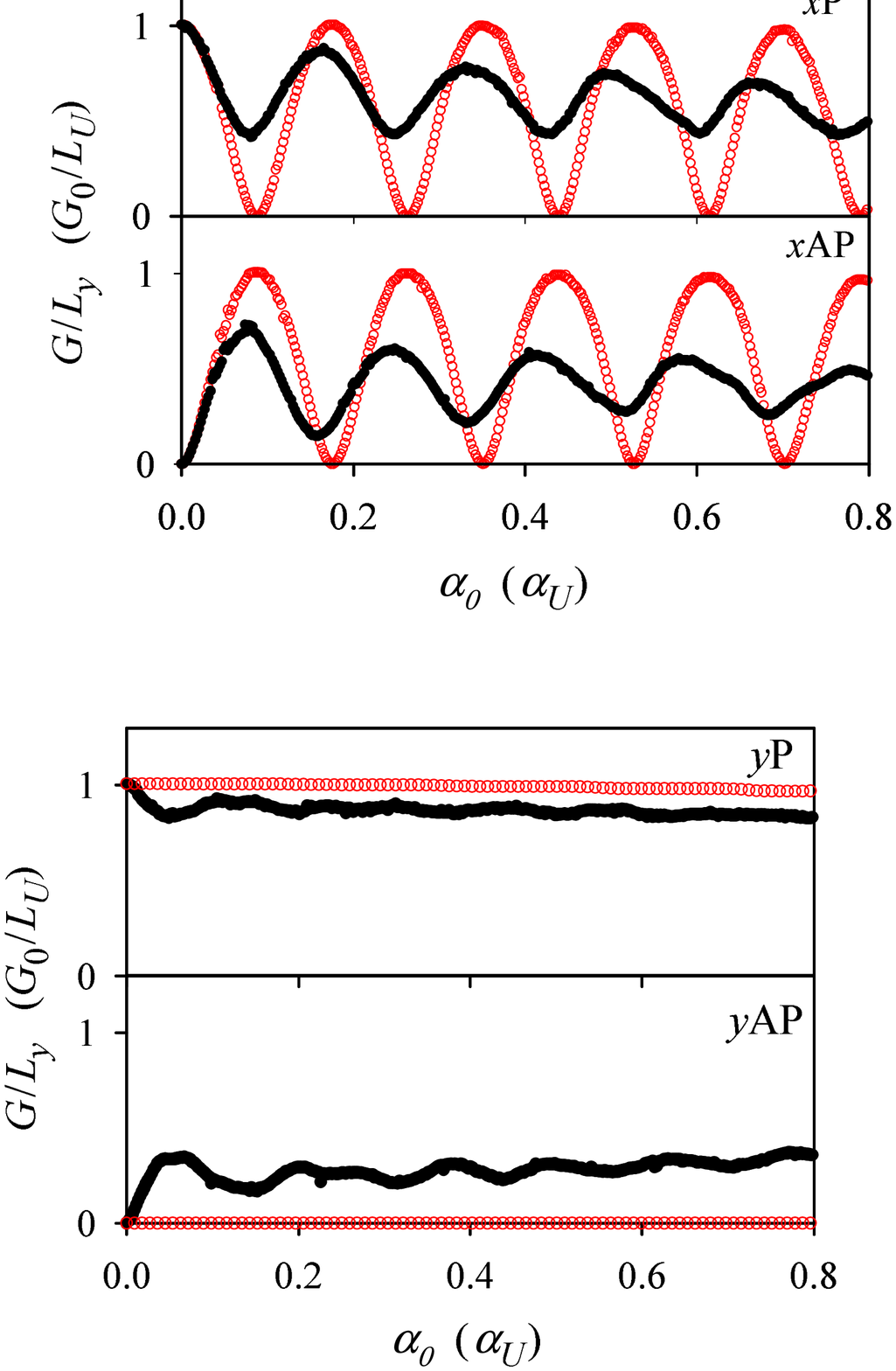,angle=0,width=0.4\textwidth,clip}
}
\caption{(Color online)
Conductance as a function of Rashba coupling intensity for 
polarization along $x$ in P and AP configurations.
Solid and open symbols are, respectively, the results with and without the 
Rashba mixing term $\alpha(x) p_y\sigma_x$ of Eq.\ (\ref{eqHR}).
Differently to the preceding figures, we take here 
the Fermi energy $E$ as energy unit, with a 
corresponding length unit $L_U=\sqrt{\hbar^2/m_0E}$. 
The Rashba-coupling unit is then $\alpha_U=\sqrt{\hbar^2 E/m_0}$.
The remaining system parameters
are: $d=20\sqrt{5}\,L_U$, $L=8\sqrt{5}\,L_U$, $\Delta_0=20\,E$.
}
\label{fig5}
\end{figure}

\begin{figure}[t]
\centerline{
\epsfig{file=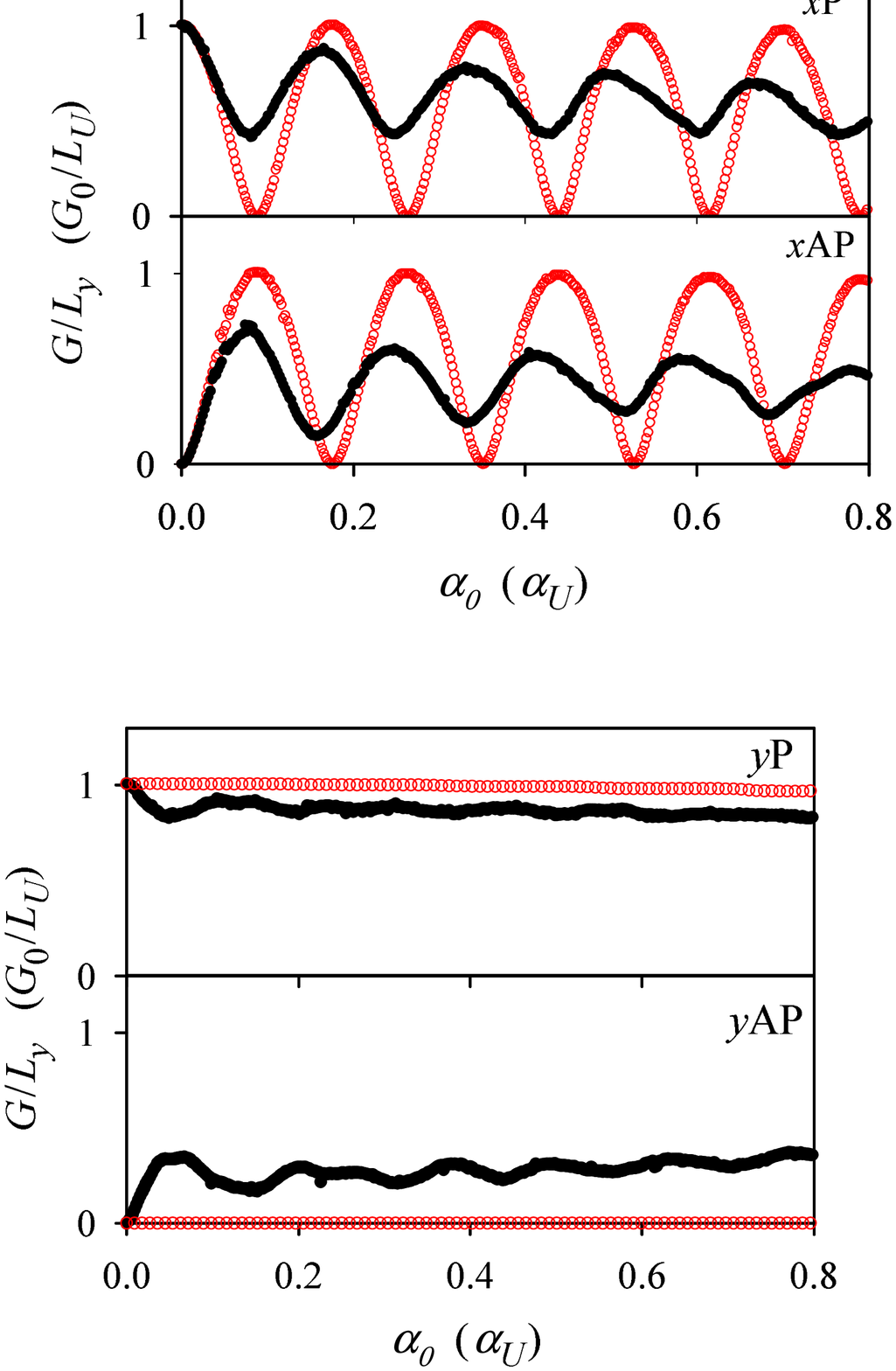,angle=0,width=0.4\textwidth,clip}
}
\caption{(Color online)
Same as Fig.\ \ref{fig5} for polarization along $y$.
}
\label{fig6}
\end{figure}

\section{Datta-Das transistor}

The Datta-Das transistor\cite{Dat90} relies on the oscillatory character of the 
conductance as a function of the Rashba intensity $\alpha_0$. In this section 
we discuss the dependence on $\alpha_0$, considering first fully
polarized contacts ($E<2\Delta_0$) and, subsequently, partial polarization 
at the end of the section.
The fully polarized results agree overall with other theoretical analysis\cite{pal04,baby,zai}
and, qualitatively,
with the experiments of Koo {\em et al.}.\cite{Koo09} Quite surprisingly, however, 
the oscillatory character of the conductance is rapidly washed out if partial
polarization is considered in our model by increasing the energy above the 
Zeeman threshold $E>2\Delta_0$.

Figure \ref{fig5} shows the results of our model for fully polarized leads
with spin oriented along $x$.
Upper and lower panels correspond, respectively, to $x$P and $x$AP configurations. 
In each case,
solid symbols represent the results for the full Rashba Hamiltonian while open
symbols correspond to the neglect of the mixing term.
The conductance shows a damped sinusoidal behavior in both cases, with  decreasing
amplitude as $\alpha_0$ rises.
These results agree with the expected Datta-Das behavior and, therefore, 
confirm the precession scenario in the continuum 2D case.
Similar damped oscillations were obtained in Refs.\ \onlinecite{pal04,baby,zai}.
Notice also that this damping is due to the Rashba mixing since it is absent
in the open symbols.
The oscillation period changes with the distance $\ell$ and successive
minima approximately fulfill the spin precession condition
$\ell\alpha_0=n\pi\hbar^2/m_0$, with $n=1,2,\dots$.
In AP configuration the 
conductance vanishes when $\alpha_0=0$ as a consequence
of the spin mismatch between both contacts, known as spin-valve behavior. 
In the presence of Rashba coupling, however, the spin valve behavior is destroyed 
and we observe that the conductance rapidly increases with $\alpha_0$, at small
couplings, and then oscillates is a similar way to the P case (Fig.\ \ref{fig5} lower panel).

For polarized leads along $z$ the results are very similar to those already discussed
for polarization along $x$ and, thus, they will not be shown. On the contrary,
Fig.\ \ref{fig6} corresponds to the configurations along $y$.
Like before, upper and lower panels are for $y$P and $y$AP configuration while
solid and open symbols represent the cases with and without band mixing, respectively.
In this case the conductance oscillations are almost absent, specially in the 
$y$P arrangement (upper panel), a result that agrees with the experiments\cite{Koo09}
and with the precessing spin scenario.\cite{Dat90}
The $y$P conductance decreases very slowly with $\alpha_0$ and the
effect of mixing is minimal, around a $10 \%$ decrease.
In $y$AP orientation we see how the Rashba mixing term again
destroys the spin-valve effect at finite $\alpha_0$'s and we observe a
small increment in conductance as the Rashba intensity increases.
In this configuration there is a reminiscence of the oscillating behavior
although much weaker as compared with the $x$ or $z$ orientations. 

\begin{figure}[t]
\centerline{
\epsfig{file=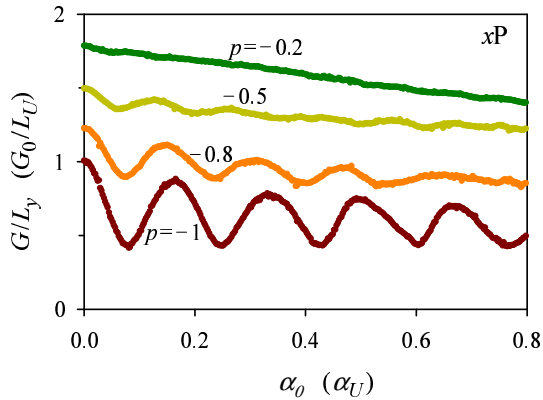,angle=0,width=0.40\textwidth,clip}
}
\caption{(Color online)
Conductance as a function of $\alpha_0$ for 
partial polarizations in $x$P configuration.
The different curves are for varying partial 
polarizations, from full ($p=-1$) to 
20\% ($p=-0.2$).
Parameters: $d=20\sqrt{5}\, L_U$, $L=8\sqrt{5}\, L_U$, units as
in Fig.\ \ref{fig5}.
}
\label{fig7}
\end{figure}

We turn now to the partially polarized cases, when the energy condition 
$E>2\Delta_0$ allows 
both spins to propagate in the contacts.
First notice that the polarization in a given 
contact $c=L,R$ is given by
\begin{equation}
p_c(E,\Delta_c)=\left\{
\begin{array}{cc}
-\displaystyle\frac{\Delta_c}{E-|\Delta_c|} & \quad (E\ge 2|\Delta_c|)\; ,\\
\rule{0cm}{0.75cm}-\displaystyle\frac{\Delta_c}{|\Delta_c|} & \quad (E\le 2|\Delta_c|)\; ,
\end{array}
\right.
\end{equation}
where, as already mentioned in Sec.\ II, we define $(\Delta_L,\Delta_R)$ to 
be $(\Delta_0,\Delta_0)$ in the P configuration and $(\Delta_0,-\Delta_0)$
in the AP configuration and $\Delta_0$ is assumed positive.
The $x$P results for partial polarizations are presented in
Fig.\ \ref{fig7}.
Notice that the oscillatory behavior is greatly quenched when the 
polarization is decreased, being heavily damped at $|p|=0.5$ and 
totally washed out at $|p|=0.2$. Thus, at low polarizations, our model 
predicts a monotonous decrease of the conductance with the intensity 
of the Rashba coupling that is not consistent with the operation of the Datta-Das
device.\cite{Dat90} 
This result shows the importance of having a high degree of polarization in the 
ferromagnetic contacts for obtaining a robust sinusoidal behavior.

The above results are not substantially modified when using other 
system parameters, 
such as changing the energy or the distance between the leads $d$.
The monotonous decrease of the conductance, without oscillations,
at low polarizations
is also seen in $x$AP, $y$P and $y$AP configurations.

\section{Space-dependent effective mass}

In this section we investigate the relevance of having
different effective masses in the semiconductor central region 
and the polarized contacts. Till now the contacts were considered
semiconductor materials with a Zeeman field in a given direction. 
A generalization towards ferromagnetic materials in the contacts 
has to include the 
different effective masses of ferromagnet and semiconductor.
As in Ref.\ \onlinecite{Mir02}, we then consider 
the effective mass in the 
contacs is
the bare electron mass $m_e$ while in the central region 
it is given by the conduction band effective mass of the 
semiconductor, $0.023m_e$ for InAs and $0.063 m_e$ for GaAs.
Our aim is not a realistic modeling of ferromagnetic contacts, 
but to explore the qualitative effects of a position 
dependent mass on the preceding full semiconductor scenario.
In particular, we shall still vary the energy from full to partial
polarization, which is not a very realistic assumption for 
a ferromagnet.

\begin{figure}[t]
\centerline{
\epsfig{file=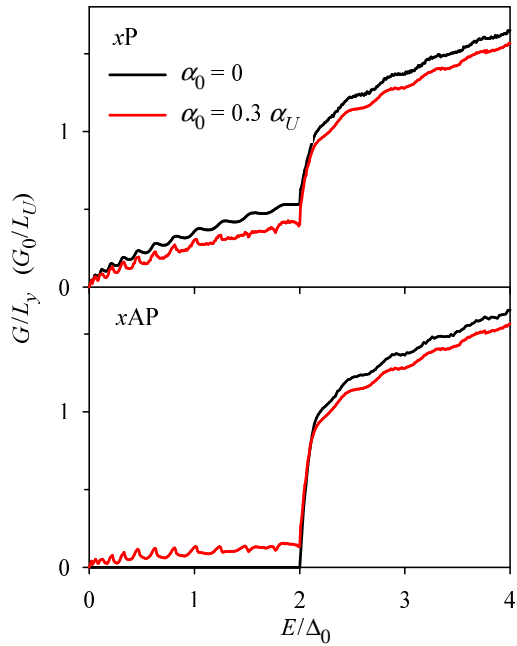,angle=0,width=0.38\textwidth,clip}
}
\caption{(Color online)
Same as Fig.\ \ref{fig2} assuming the effective mass in the contacts
is given by $m_c=15 m_0$ ($c=L,R$).}
\label{fig8}
\end{figure}

In the spirit of the model, we now use a 
generalized kinetic term with a position dependent effective 
mass $m(x)$ (see Appendix \ref{appA}) evolving from the semiconductor
mass $m_0$ in the central region to $15m_0$ in both contacts; 
\begin{equation}
T_{\rm kin}=
-\frac{d}{dx}\frac{\hbar^2}{2m(x)}\frac{d}{dx}
-\frac{\hbar^2}{2m(x)}\frac{d^2}{dy^2}\, .
\end{equation}
The big jump in effective mass at the interface is 
smoothed using Fermi functions as explained in Appendix \ref{appA}.
The presence of these {\em effective-mass interfaces} is an additional source
of conductance oscillations, as compared to the discussion of the 
preceding sections. Indeed, in this case even the $\alpha_0=0$ conductance
with fully polarized contacts displays Ramsauer oscillations, as shown
by the black symbols in the upper panel of Fig.\ \ref{fig8}.
Based on the preceding section results, the addition of the spin orbit
coupling is expected to introduce new oscillations  of Fano
type due to the 
coupling with quasibound states.
Surprisingly, both types of oscillations interfere destructively,
specially in the vicinity of the polarization 
threshold $E=2\Delta_0$, as shown by the gray (red) symbols in Fig.\ \ref{fig8} upper panel.
Another conspicuous effect of the effective mass discontinuity is the 
big enhancement of conductance when the energy exceeds $2\Delta_0$. This 
is clearly noticed when comparing the 
upper panels of Figs\ \ref{fig8} and \ref{fig2}.

The lower panel of Fig.\ \ref{fig8} shows the $x$AP conductance with 
position dependent effective mass. As a difference with Fig.\ \ref{fig2}
(lower panel),
there are Ramsauer
oscillations due to the mass jumps for $E>2\Delta_0$
even for a vanishing $\alpha_0$. Below threshold we find Rashba-induced 
oscillations that look very similar to the $x$P ones in the upper panel.
For $y$ oriented contacts the results (not shown), as compared to those 
of Fig.\ \ref{fig2b}, are also characterized by 
the appearance of clear Ramsauer oscillations below threshold
in the $y$P configuration while in the $y$AP orientation the variations are much 
smaller.

\begin{figure}[t]
\centerline{
\epsfig{file=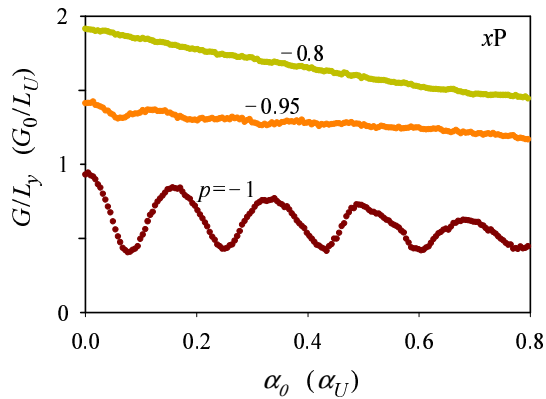,angle=0,width=0.4\textwidth,clip}
}
\caption{(Color online)
Same as Fig.\ \ref{fig7} assuming the effective mass in the contacts
is given by $m_c=15 m_0$ ($c=L,R$).
}
\label{fig9}
\end{figure}

A natural question to ask is whether the effective mass modification
affects the conductance oscillations with $\alpha_0$ discussed
in Sec.\ IV. This is addressed in Fig.\ \ref{fig9}
for the fully polarized $x$P configuration. There are small changes,
of course, but the overall behavior with damped oscillations is well 
preserved. Another result we should check is the disappearance of
the oscillations at partial polarizations of the contacts (Fig.\ \ref{fig7}).
As proved by Fig.\ \ref{fig9}, this result is also robust with 
respect to effective mass changes. Actually, the quenching of the
oscillations at partial polarization is enhanced when the mass
in the contacts is taken to be the bare mass: already for $|p|<0.8$ the
conductance becomes monotonous, having only a slight decrease with 
$\alpha_0$.

\section{Conclusions}

The linear conductance of a 2D stripe of Rashba interaction with 
polarized contacts behaves in characteristic ways depending on the 
electron energy, the Rashba intensity as well as on the 
spin orientation and the degree of polarization of the contacts.
We have explored all these degrees of freedom using a single model 
based on an effective mass approach combined with
phenomenological Rashba coupling and Zeeman fields. Our analysis was mostly
numerical, in order to avoid additional simplifying assumptions.
Nevertheless, semianalytical analysis in terms of coupled channels and
quasibound states was also provided.

We have identified two types of oscillations: Ramsauer oscillations
due to discontinuities in the effective potential and effective mass;
and Fano oscillations due to the coupling with quasibound states.
The latter ones are exclusively due to the Rashba coupling.
For energies below the full polarization threshold and 
without mass jumps at the interfaces we obtained
pure Fano oscillations in both parallel and antiparallel polarizations
along $x$, while they are absent for polarization along $y$. 
With the addition of mass jumps at the interfaces the
oscillations are quenched; a result that we attribute to
a destructive interference between Fano and Ramsauer oscillations.
Above the polarization threshold the conductance displays Ramsauer
oscillations in most cases.

Regarding the oscillations in conductance  as a function of 
Rashba coupling, our main result is the rapid quenching of the 
oscillations when the contacts are partially polarized. 
This oscillation quenching is even more pronounced when
the effective mass increases in the contacts. 
The oscillations disappear for polarizations below
20\% and 80\% for constant and position-dependent effective mass,
respectively.

\section*{Acknowledgments}
This work was supported by the MICINN (Spain) Grant FIS2008-00781.
Useful discussions with D. S\'anchez and R. L\'opez are gratefully 
acknowledged.

\appendix

\section{Smooth transitions}
\label{appA}

This appendix contains the precise mathematical forms of the smooth functions
$\alpha(x)$, $\Delta(x)$ and $m(x)$ defining the system Hamiltonian.
We model the step-like character of these quantities using Fermi functions
\begin{equation}
{\cal F}_{x_0,\sigma}(x)=
\frac{1}{1+e^{(x-x_0)/\sigma}}\; ,
\end{equation}
where $x_0$ is the position of the step and $\sigma$ is giving the length around
$x_0$ in which the transition takes place. Precisely, it is
\begin{eqnarray}
\label{A1s2}
\alpha(x) &=& 
\alpha_0\, \left[{\cal F}_{\ell/2,\sigma}(x)-{\cal F}_{-\ell/2,\sigma}(x)\right]\; ,\\
\label{A1s3}
\Delta(x) &=& 
\Delta_L {\cal F}_{-d/2,\sigma}(x)
+
\Delta_R\, \left[ 1-{\cal F}_{d/2,\sigma}(x) \right]\,,\\
\label{A1s4}
m(x) &=& 
m_L {\cal F}_{-d/2,\sigma}(x)
+
m_R\, \left[1-{\cal F}_{d/2,\sigma}(x)\right]\nonumber\\
&+&
m_0 \, \left[ {\cal F}_{d/2,\sigma}(x)-{\cal F}_{-d/2,\sigma}(x)\right]\; .
\end{eqnarray}
where the constants are: $\alpha_0$, Rashba intensity; 
$\Delta_L$, $\Delta_R$, Zeeman fields;
$m_L$, $m_0$ and $m_R$, effective masses.
Notice that, for the sake of simplicity, we assume a common 
value for $\sigma$ in 
Eqs.\ (\ref{A1s2}), (\ref{A1s3}) and (\ref{A1s4}).

\section{Resolution method}
\label{appB}

The numerical calculation of the linear conductance Eq.\ (\ref{eqG}) at 
a given energy $E$ involves two steps. First, for a certain angle $\theta$,
or what is equivalent, a certain transverse momentum $q$, the coupled equations 
for $\psi_{q+}$ and $\psi_{q-}$, Eqs.\  (\ref{ccmeq}), are solved to obtain the 
transmissions
$T_{s's}$ and $T'_{s's}$. This is accomplished 
using the transmitting-boundary algorithm as in Ref.\ \onlinecite{gel10}.
The calculation of Ref.\ \onlinecite{gel10} was for quantum wires, with a 
confinement potential in the transverse direction, where the system of
coupled equations was infinite and had to be truncated. The present
case is, in this respect, simpler since only the two spin components 
of a given transverse momentum need to be considered. Nevertheless, the reader is addressed
to Ref.\ \onlinecite{gel10} for the technical details on how the differential
equations with open boundary conditions are transformed into a linear system
of equations.\cite{HSL}

Once the transmissions at a fixed $\theta$ are obtained, a second step
of the calculation requires to integrate over the angle to calculate the linear conductance from 
Eq.\ (\ref{eqG}).
This integral turns out to be somewhat delicate due to the presence of resonances
as discussed in the quasi-analytical solution by ansatz of Sec.\ III.B. The
$\theta$-integration is then carried out using Gauss-Legendre quadratures with a certain
set of abcissae and weights. To make sure that the integral is well converged we
keep increasing the number of Gauss-Legendre points until a required accuracy 
is reached in a stable way. Typically, we require the error to be $\Delta G/G_0\leq 10^{-3}$.

\end{document}